\documentclass[aps,prb,twocolumn,superscriptaddress,citeautoscript]{revtex4-1}

\usepackage{amsmath,amssymb}
\usepackage{bm}
\usepackage{hyperref} %for links in pdf

\begin{document}

\title{Energy density of variational states }

\author{Leon Balents}
\affiliation{Kavli Institute for Theoretical Physics, University of
  California, Santa Barbara, CA 93106-4030, U.S.A.}

\date{\today}
\begin{abstract}
  We show, in several important and general cases, that a low variational energy {\em density} of a trial state is possible even when the trial state represents a different phase from the ground state.  Specifically, we ask whether the ground state energy density of a Hamiltonian whose ground state is in phase A can be approximated to arbitrary accuracy by a wavefunction which represents a different phase B.  We show this is indeed the case when A has discrete symmetry breaking order in one dimension or topological order in two dimensions, while B is disordered.  We argue that, if reasonable conditions of physicality are imposed upon the trial wavefunction, then this is {\em not} possible when A has discrete symmetry breaking in dimensions greater than one and B is symmetric, or when A is topologically trivial and B has topological order.  \end{abstract}


\maketitle

\section{Introduction}
\label{sec:introduction}

Quantum many body theory has revealed the existence of ``exotic'' ground states phases of matter -- which we take to mean phases distinguished by means other than symmetry breaking.  Notable are topological phases, which are fully gapped ground states supporting anyonic excitations in two dimensions, and embodying a generalization of this structure in three dimensions.  A topological phase is locally stable to {\em all} perturbations, making it apparently more robust even than a conventional symmetry breaking state.  While the theoretical existence of a huge family of topological phases is well-established through exactly soluble models and an extensive formalism, these states are so far scarce in the real world.  

The identification of these exotic phases is hampered by the dual challenges of calculating ground states of many body systems, and of identifying the hidden topological or other structure within those states.  Consequently, a prominent computational approach is the variational method, in which a trial wavefunction is optimized within a given (say) topological class, and the energy of such optimized wavefunctions can be compared across classes.  A popular example is the use of Gutzwiller projected free fermion states for quantum spin systems. In this context exotic phases are known as {\em quantum spin liquids}.   Distinct states can be constructed using the projective symmetry group construction\cite{wen2002quantum}, and the energy of optimized states compared.  A natural implication, presumed in many studies,\cite{iqbal2013gapless,lu2011z,clark2011nature,hermele2008properties,zhou20084,lawler2008gapless,yunoki2006two,motrunich2005variational,ran2007projected} is that the true ground state is in the same phase as the wavefunction with lowest optimized energy.  

In this paper, we discuss the reliability of this conclusion for systems with different types of ground states.  We consider {\em local} Hamiltonians $H$ for {\em infinite} quantum systems of dimensionality $d=1,2,3$.  For finite systems the variational method is on very firm ground: if a state $|\psi\rangle$ has variational energy $E_v = \langle \psi|H|\psi\rangle$ which is separated by the ground state by an amount less than the spectral gap, $E_0<E_v<E_1$, where $E_0$ is the exact ground state energy and $E_1$ is the first excited state energy, then $|\psi\rangle$ {\em must} have a finite overlap with the exact ground state, and the minimum value of this overlap increases as $E_v$ approaches $E_0$.  However, for infinite systems, there is a fundamental difficulty: both the variational energy and the exact ground state energy are {\em infinite}, and generically, $E_v$ and $E_0$ differ by an amount proportional to the system volume $L^d$.  Standard practice is to compare the variational {\em energy density} (e.g. per site or per unit volume) of different states.  The variational principle guarantees that the variational energy density is bounded below by the exact energy density, so one can certainly use the variational method to get a good estimate of the ground state energy density.  But is it actually predictive for the {\em phase} of the ground state?

Many studies of strongly correlated problems (e.g. Hubbard models, frustrated quantum magnets) have shown that states belonging to different phases have very close variational energy densities.  A common interpretation of this observation is that the actual system is somehow indecisive about its true ground state, i.e. that several phases are ``close'' in phase space, and could be selected in the true ground state by small perturbations of the original Hamiltonian.  This notion is somehow similar to the picture of first order phase transitions, for which near the transition the higher (free) energy phase exists as a metastable object even when it is not the global thermodynamic minimum.  According to this line of thinking, distinct variational wavefunctions represent ``competing orders'' that play a central role in the physics.  However, it is not a priori obvious that the phase of the variational state has predictive value.

This paper addresses this issue at the level of principle.  Specifically, we ask whether a wavefunction in the {\em wrong} phase can yield an {\em arbitrarily} good approximation to the ground state energy density for a given Hamiltonian.  It is natural to suppose that the answer to this question is the same for all non-fine-tuned local Hamiltonians in a particular dimension, whose ground state is in a given phase.  Then we say that a given phase A in a given dimension is {\em variationally robust} to phase B, if for all non-fine-tuned Hamiltonians with a ground state in that phase in that dimension, states which belong to a different phase B have variational energy density which is strictly larger than that of the ground state.  Conversely, if variational states in phase B can be found with energy density arbitrarily close to the exact ground state value, then we say that the phase A is {\em not} variationally robust to phase B.  Clearly variational methods are more predictive for states that are variationally robust.  

Theoretically, it is much easier to show that a state is {\em not} variationally robust, simply by providing a constructive example of a variational state in the wrong phase with low energy density.  Demonstrating variational robustness is much more difficult, and indeed we do not know how to {\em prove} this rigorously.  Part of the difficulty is that one must specify somehow the allowed set of possible variational states.  We would like to include only states that are characteristic of generic ground state phases of local Hamiltonians.  This is a small subset of all possible quantum states.  For example, any state which exhibits a volume law of entanglement entropy cannot be a valid ground state. 

With this in mind, we restrict ourselves to {\em gapped} phases of bosonic/spin systems, for which it is generally agreed that ground states can be described by tensor network states (TNS), also know as projected entangled pair states (PEPS).  We therefore assume that {\em both} the exact ground state and the variational state have the TNS form. An important fact is that the variational energy density, being an expectation value of local operators, is a smooth function of the tensor describing the TNS.  Then one sees that the question of variational robustness is equivalent to the robustness of the phase in question with respect to small variations of the TNS tensor.  If an infinitesimal change of the tensor changes the phase of the TNS describing the exact ground state, then the phase is {\em not} variationally robust.  Conversely, a variationally robust phase must be preserved by {\em all} infinitesimal variations of the TNS tensor.  If the latter condition is true, the phase is likely to be variationally robust, though this is not a rigorous proof.

The TNS description applies only to fully gapped phases.  Thus we can consider phases with no order, with discrete symmetry breaking, and with topological order.  We first consider the situation in which ground state has no order, and show that broken symmetry states with arbitrarily low variational energy {\em always} exist.   This is very straightforward and indeed agrees with a simplistic picture based on Landau theory.  Next, we consider the converse situation, and ask whether a broken symmetry phase is variationally robust: do low energy variational states exist in which the symmetry in question is {\em unbroken}?  We show that discrete broken symmetry is {\em not} variationally robust in one dimension, but argue that it {\em is} robust in two or more dimensions.  We then turn to topological order, which exists only in dimensions two or greater.  We show that topological order is {\em not} variationally robust in two dimensions, i.e. that low energy states without topological order exist even when the ground state has topological order.  This shows that topological order is less robust in the variational sense than discrete symmetry breaking.  We do not make general statements about topological phases in three dimensions, but do discuss the case of the toric code ($Z_2$ topological phase).  Finally, we consider the converse question: whether a phase without topological order admits low energy variational states {\em with} topological order.  We argue this is not possible, so that topologically trivial phases are variationally robust to approximants with topological order.

The remainder of this paper discusses these ideas in more detail. Sec.~\ref{sec:discr-symm-break-1} discusses symmetry breaking order, and robustness of states with or without such order to those without/with broken symmetry.  ones.  Next, in Sec.~\ref{sec:topol-order-two}, we discuss topological order, and the robustness of states with or without such order.  Finally, we conclude with some general comments in Sec.~\ref{sec:discussion}.

\section{Discrete symmetry breaking}
\label{sec:discr-symm-break-1}

\subsection{Robustness of ``disordered'' states to symmetry breaking}
\label{sec:robustn-disord-stat}

Here we consider a Hamiltonian which is invariant under a discrete symmetry group $\mathcal{G}$.  Suppose there exists a local order parameter $\phi$ which is not invariant under $\mathcal{G}$ ($\phi$ may be a scalar or have multiple components).   Suppose that the ground state $|0\rangle$ is invariant under $\mathcal{G}$.  We ask whether a variational state $|\phi\rangle$ exists with non-zero $\phi$ and arbitrarily low variational energy density.   A simplistic Landau theory arguments suggests yes.  We write the energy density $\mathcal{E}$ for the state $|\phi\rangle$ as a function of $\phi$.  Standard Landau reasoning suggests that it takes the form
\begin{equation}
  \label{eq:17}
  {\mathcal{E}}(\phi) = \mathcal{E}_0 + r \phi \cdot \phi + \cdots,
\end{equation}
where $\phi \cdot \phi$ is the quadratic invariant of the order parameter, and $r>0$ is required if we assume the ground state has $\phi=0$ and unbroken symmetry.  Clearly this predicts that a broken symmetry state with arbitrarily low energy density can be found simply by taking the magnitude of $\phi$ small but non-zero.  

Microscopically, we can argue similarly, by the following construction.  We assume the Hamiltonian $H$ has the symmetry $\mathcal{G}$ and ground state $|0\rangle$.  Consider a fiduciary Hamiltonian $H_\lambda$, with
\begin{equation}
  \label{eq:18}
  H_\lambda = H - \lambda \int d^dx\, \hat\phi,
\end{equation}
where the integral over space can be replaced by a sum, as appropriate, and $\hat\phi$ is a local operator with the same symmetry as the order parameter.  The ground state of $H_\lambda$ smoothly goes to that of $H$ as $\lambda\rightarrow 0$, and hence so does the energy density.  We can consider this state, $|\lambda\rangle$ as a variational state for $H$.  Then
\begin{equation}
  \label{eq:19}
  \langle\lambda |H|\lambda\rangle = {\mathcal{E}}_0(\lambda) + \langle\lambda |\hat\phi|\lambda\rangle,
\end{equation}
where $\mathcal{E}_0(\lambda)$ is the ground state energy of $H_\lambda$.  By perturbation theory (which is valid because, by assumption, $H$ has a gap), we see that the right hand side above is equal to $\mathcal{E}_0$ up to terms of order $\lambda^2$.  Hence the state $|\lambda\rangle$, which has broken symmetry whenever $\lambda\neq 0$, has a variational energy density of order $\lambda^2$ higher than the exact ground state.

So we conclude that a disordered (or symmetric) phase is {\em never} variationally robust, in any dimension, to a symmetry-broken state.  This is not surprising and agrees with the intuition based on simple Landau theory.

\subsection{Robustness of symmetry broken phases to approximation by symmetric states}
\label{sec:discr-symm-break}

Now we consider the reverse question from that of the previous section.  Suppose the ground state of $H$ breaks a discrete symmetry.  Can one find a state with {\em unbroken} symmetry with arbitrarily low variational energy density?  Here Landau theory suggests the answer is no.  This is because the symmetry broken minima in the Landau energy functional are separated by a non-zero distance from the symmetric point: a small variation of a non-zero order parameter $\phi$ cannot restore the symmetry.  This argument is, however, na\"ive, and demonstrably incorrect in one dimension, as we now show.

\subsubsection{Example: Transverse field Ising chain}
\label{sec:transv-field-ising}

Let us begin with the archetypal example of symmetry breaking in one dimension, the transverse field Ising model,
\begin{equation}
  \label{eq:1}
  H = \sum_i - \sigma_i^z \sigma_{i+1}^z - h \sigma_i^x,
\end{equation}
where $\vec\sigma_i$ are Pauli matrices.  The model is exactly soluble, and, for $|h|<1$, has a ferromagnetic ground state which breaks the Ising symmetry, $\sigma_i^z \rightarrow - \sigma_i^z$.  There is an excitation gap above the ground state, and exponentially decaying connected correlations.  Consequently, the ground state can be well-approximated by a matrix product state,
\begin{equation}
  \label{eq:2}
  |\psi\rangle = \sum_{\{\sigma_i = \pm 1\}} {\rm Tr}\, \left[ {\sf M}(\sigma_1){\sf M}(\sigma_2)\cdots {\sf M}(\sigma_N)\right] |\sigma_1\cdots \sigma_N\rangle, 
\end{equation}
where $\sigma_i=\pm 1$ denotes the eigenvalue of $\sigma_i^z$.  A broken symmetry ground state with positive magnetization is described by a matrix ${\sf M}_+(\sigma)$ which preferentially favors $\sigma=+1$ over $\sigma=-1$.  One can obtain the other ground state by transforming the matrix into ${\sf M}_-(\sigma) = {\sf M}_+(-\sigma)$.  For a finite system, one expects exponentially weak tunneling to mix the two symmetry broken states, so that the proper system eigenstates are the Schr\"odinger cat states $|{\sf M}\rangle$ defined by the enlarged matrix 
\begin{equation}
  \label{eq:4}
  {\sf M} =
  \begin{pmatrix}
    {\sf M}_+ & 0 \\ 0 & e^{i\phi}{\sf M}_- 
  \end{pmatrix},
\end{equation}
where we may choose $\phi = 0$ and $\phi = \pi/N$ to obtain linearly independent states, even and odd linear combinations of the two ferromagnetic states.  The energy {\sl density} of the Schr\"odinger cat states is identical to that of pure ferromagnetic states up to exponential corrections, and indeed they represents the true eigenstates of large finite systems.  Moreover these states still have broken symmetry in the sense of off-diagonal long-range order: the correlation functions obey 
\begin{equation}
  \label{eq:6}
  \langle {\sf M}| \sigma_i^z \sigma_j^z |{\sf M}\rangle \rightarrow \overline\sigma^2 \qquad \mbox{for } 1\ll |i-j| \ll N.
\end{equation}

Now we claim that an {\em infinitesimal} deformation of the matrix in Eq.~\eqref{eq:4} leads to a state with {\em unbroken} Ising symmetry.  Specifically, consider
\begin{equation}
  \label{eq:5}
   {\sf M}_\epsilon =
  \begin{pmatrix}
    {\sf M}_+ & \epsilon{\sf M}_1 \\ \epsilon {\sf M}_2 & {\sf  M}_- 
  \end{pmatrix}.
\end{equation}
Provided ${\sf M}_1(-\sigma) = {\sf M}_2(\sigma)$, the state $|{\sf M}_\epsilon\rangle$ retains the Ising symmetry, as can be easily verified.  However, when $\epsilon \neq 0$, the long-range order of the state is destroyed.  This is true in general, but we illustrate it for transparency for the simplest case in which ${\sf M}_\pm$ are one dimensional, which corresponds for $\epsilon=0$ to a mean field approximation to the ground state.

To see the destruction of long-range order, we can directly calculate the spin-spin correlation function.  Following standard methods, we have
\begin{equation}
  \label{eq:7}
  \langle {\sf M}_\epsilon| \sigma_i^z \sigma_j^z |{\sf M}_\epsilon\rangle = \frac{1}{Z}{\rm Tr} \left[ {\sf T}^{N-|i-j|-1} {\sf S} {\sf T}^{|i-j|-1} {\sf S}\right],
\end{equation}
where the trace is expressed in a doubled Hilbert space which is defined as a direct product of the original matrix product one, owing to the product of two wavefunctions in the expectation value.  We have
\begin{eqnarray}
  \label{eq:8}
  {\sf T} & = & \sum_{\sigma} {\sf M}_\epsilon(\sigma) \otimes {\sf  M}_\epsilon(\sigma), \\
  {\sf S} & = & \sum_{\sigma} \sigma{\sf M}_\epsilon(\sigma) \otimes {\sf  M}_\epsilon(\sigma),
\end{eqnarray}
and $Z = {\rm Tr}\, T^N$.  For the one dimensional case, we have ${\sf M}_\pm(\sigma) = \sqrt{(1\pm m \sigma)/2}$, where $m$ is the spontaneous magnetization when $\epsilon=0$, and $-1<m<1$.  We take ${\sf M}_1={\sf M}_2=1$. Expressing the matrices in the direct product space in terms of two sets of Pauli matrices $\vec\tau_1$ and $\vec\tau_2$, we have then
\begin{eqnarray}
  \label{eq:9}
  {\sf T} & = & 2 \left[ (m_+ + \epsilon \tau_1^x) (m_+ + \epsilon \tau_2^x) + m_-^2 \tau_1^z \tau_2^z\right], \\
  {\sf S} & = & 2 \left[m_+ m_- (\tau_1^z+\tau_2^z) + m_- \epsilon (\tau_1^z \tau_2^x + \tau_1^x \tau_2^z)\right],
\end{eqnarray}
where 
\begin{equation}
  \label{eq:10}
  m_\pm = \frac{1}{2} \left( \sqrt{\frac{1+m}{2}}\pm \sqrt{\frac{1-m}{2}}\right).
\end{equation}
Expressing Eq.~\eqref{eq:7} in terms of the eigenstates of ${\sf T}$ one obtains
\begin{equation}
  \label{eq:11}
  \langle {\sf M}_\epsilon| \sigma_i^z \sigma_j^z |{\sf M}_\epsilon\rangle = \frac{\sum_{a,b}  t_a^N \left(\frac{t_b}{t_a}\right)^{|i-j|-1} |\langle a|{\sf S}|b\rangle|^2}{\sum_a t_a^N},
\end{equation}
where ${\sf T}|a\rangle = t_a |a\rangle$ and $\langle a|a\rangle=1$, and we can choose a convention in which $t_0 \geq t_1 \geq t_2 \geq t_3$.  In the thermodynamic limit, $N\rightarrow\infty$, this is dominated by the term or terms in which $t_a$ are maximal.  For $\epsilon\neq 0$, $t_0>t_1$, and the maximal eigenvalue is unique, and so
\begin{equation}
  \label{eq:12}
   \langle {\sf M}_\epsilon| \sigma_i^z \sigma_j^z |{\sf M}_\epsilon\rangle \xrightarrow [N\rightarrow\infty]{\epsilon\neq0}  \sum_{b}  \left(\frac{t_b}{t_0}\right)^{|i-j|-1} |\langle 0|{\sf S}|b\rangle|^2.
\end{equation}
By Ising symmetry, one readily sees that $\langle 0|{\sf S}|0\rangle=0$, so all terms in the sum decay exponentially (this breaks down for $\epsilon=0$ because the leading eigenvalues are degenerate, $t_1=t_0$ in that case), and there is no long-range order.  For $\epsilon \ll 1$, we can approximate $|\langle 0|{\sf S}|1\rangle|^2 \approx m^2$, $t_b/t_0 \approx 1 - 8 \epsilon^2/(1-\sqrt{1-m^2})$, so that
\begin{equation}
  \label{eq:13}
  \langle {\sf M}_\epsilon| \sigma_i^z \sigma_j^z |{\sf M}_\epsilon\rangle \xrightarrow [N\rightarrow\infty]{0<\epsilon\ll 1} m^2 e^{-|i-j|/\xi},
\end{equation}
with 
\begin{equation}
  \label{eq:14}
  \xi \approx \frac{1-\sqrt{1-m^2}}{8 \epsilon^2}.
\end{equation}
So we see that the state $|{\sf M}_\epsilon\rangle$ has, for arbitrarily small $\epsilon$, no long-range order.  

\subsubsection{Domain wall interpretation}
\label{sec:doma-wall-interpr}

For the toy model in the previous subsection, it is straightforward to interpret the physics of the variational state.  In general, the two blocks of the double matrix in Eqs.~(\ref{eq:4},\ref{eq:5}) describe the two physical uniform Ising ground states.  By introducing a small non-zero $\epsilon$ in Eq.~\eqref{eq:5}, we introduce a probability amplitude $\epsilon$ for a {\em domain wall} to appear in this uniform state.  A domain wall is a topological soliton in the Ising ordered phase, which in one dimension is a localized point-like defect separating the two Ising domains.  It is an {\em excitation} of the Ising ordered phase (indeed it is the elementary one).  In general, in a discrete broken symmetry state, a non-zero gap is expected for such an excitation.  The state $|{\sf M}_\epsilon\rangle$ can be viewed as a state with a superposition of domain walls, whose mean density becomes small for $\epsilon \ll 1$ .  The typical distance between domain walls is just the length $\xi$ above.  If the correlation function is measured between two points whose separation is much larger than $\xi$, the number of domain walls between these two points fluctuates randomly (or more properly the wavefunction is a superposition of components in which this number varies significantly), leading to contributions to the correlation function of opposite sign that cancel and decay exponentially with length.

In the {\em variational} sense, a low density superposition of domain walls has a low energy density in the Ising ordered phase, since the energy of each soliton is finite, and the number per unit length is $1/\xi$.  Hence the variational energy density of such a state above the ground state is expected to be of order $\Delta/\xi$, where $\Delta$ is the excitation gap for a single soliton.  This can be made arbitrarily small by increasing $\xi$.  

Some comments are in order.  The domain wall state is completely spatially uniform.  It is also {\em not} a thermal state. Since it can be expressed as a matrix product state, it clearly obeys the area law for entanglement entropy, unlike a thermal state.   Hence it can be a plausible ground state for some locally interacting quantum system.  In fact, it is natural to regard the state as an approximation to the ground state in the vicinity of the quantum critical point of the Ising chain, on the quantum paramagnetic side.  It is a quasi-condensate of the solitons, if we view the latter as particles.  Note of course it would be a crude approximation to the actual ground state near the critical point of the chain, and does not capture the universal physics of the quantum critical regime.  It does, however, roughly capture the universal physics of the quantum paramagnetic phase.  In this paper, we wish to stress only that in this way one constructs a wavefunction which describes a symmetry unbroken state, which could be a good approximation to the ground state of some local Hamiltonian, and which approximates the ground state energy of the {\em ferromagnet} arbitrarily well.  Thus we have shown, by contradiction, that the statement that a good variational energy density is a predictor for ferromagnetism is false in one dimension.

From this physical picture, it is clear that the same must be true for {\em all} discrete symmetry breaking states in one dimension.  The elementary topological excitations for all such states are pointlike solitons, and the proliferation of such solitons destroys the broken symmetry phase.  Since they are local, these solitons have finite energy, and so states of arbitrarily small energy density can be constructed by making the solitons dilute.  

We return to the connection to thermal fluctuations.  It is well-known that one dimensional systems at non-zero temperature $T>0$ cannot display spontaneous symmetry breaking of discrete symmetries.  This is because soliton excitations cost a finite energy, and so appear with non-zero probability per unit length by thermal activation, so over a sufficiently large length, many solitons inevitably appear, and long-range order is destroyed.  This is {\em not} the case at $T=0$.  The presence of a finite energy gap {\em is} sufficient to protect the Ising ordered phase at $T=0$.  However, the expectation value of physical observables within a matrix product state can generically be expressed as a local classical statistical mechanics model in the same dimension, in which both the physical spins and the auxiliary ones (matrix degrees of freedom) appear as fluctuating variables.  In the physical ferromagnetic ground state, the corresponding classical statistical mechanics problem has a zero temperature character.  Thus is can sustain long-range order.  However, when perturbed with non-zero $\epsilon$, the fictitious statistical mechanics problem develops a non-zero small fictitious temperature.  Consequently, the classical reasoning applies to this fiduciary problem, and we may understand the destruction of long-range order by non-zero $\epsilon$ by the corresponding destruction at $T>0$ in the physical case.  

\subsubsection{Robustness of symmetry broken states in $d \geq 2$}
\label{sec:robustn-symm-brok}

We now consider the case of $d \geq 2$.    In analogy to the approach of Sec.~\ref{sec:transv-field-ising}, we consider a TNS or PEPS representation of the ground state, which generalizes naturally the MPS to $d>2$.  For a spin system, a tensor is defined for every site containing a spin, with one tensor index for each link connecting that site to another neighboring site.  The link indices, which generalize the matrix indices in the MPS, comprise fiduciary variables to be summed over in building the wavefunction:
\begin{equation}
  \label{eq:20}
  |\psi\rangle = \sum_{\{\sigma_i\} } \sum_{abc\cdots} \left[ T_{abcd}(\sigma_1) T_{aefg}(\sigma_2)\cdots \right] |\sigma_1\cdots\sigma_N\rangle.
\end{equation}
Here $a,b,c,\cdots$ indicate the link indices, which are summed over $D$ values, defining the ``inner dimension'' of the TNS, and $T_{abcd}(\sigma_1)$ is the tensor at site $1$ etc. (the choice of 4 links per site was arbitrary).  Each link index occurs in just two tensors, and the arrangement of links defines the network.  

It is straightforward to write down tensors corresponding to symmetry broken states.  For example, the cat state of Eqs.~\eqref{eq:4} in the simplest case $D=2$ is 
\begin{equation}
  \label{eq:21}
  T^{(\epsilon)}_{abcd}(\sigma)  =  \left\{\begin{array}{cc}
      \sqrt{\frac{1+a\sigma m}{2}} & a=b=c=d, \\
      \mathcal{O}(\epsilon) & \textrm{otherwise}
      \end{array}\right. ,
  \end{equation}
where we take the inner indices $a=\pm 1$, and $m>0$ is the magnetization.  If $\epsilon=0$, this is a perfect cat state, but one can include defects in the cat state (analogous to the domain walls in 1d) by allowing non-zero off-diagonal entries, $\epsilon >0$.  

The behavior of the corresponding TNS, however, is very different from that of the 1d MPS.  In fact, the long-range order of the TNS is robust to {\em arbitrary} small perturbations.  This can be most simply understood from the mapping of TNS matrix elements to classical statistical mechanics.  Specifically, the spin-spin expectation value
\begin{equation}
  \label{eq:22}
  C_{ij} = \langle \psi| \sigma^z_i \sigma^z_j|\psi\rangle/\langle \psi|\psi\rangle,
\end{equation}
which diagnoses long-range order ($C_{ij} \rightarrow C_\infty \neq 0$ as $|i-j|\rightarrow \infty$) can be expressed via Eq.~\eqref{eq:20} as
\begin{equation}
  \label{eq:23}
  C_{ij} = \frac{\sum_{\{ \sigma_i \}}\sum_{\stackrel{ab\cdots}{\scriptscriptstyle a'b'\cdots}} \sigma_i \sigma_j W[\{ \sigma_i\}, abc\dots, a'b'c'\cdots]}{\sum_{\{ \sigma_i \}}\sum_{\stackrel{ab\cdots}{\scriptscriptstyle a'b'\cdots}}W},
\end{equation}
where
\begin{eqnarray}
  \label{eq:24}
&&  W[\{ \sigma_i\}, abc\dots, a'b'c'\cdots]  =  \left[ T_{abcd}(\sigma_1) T_{aefg}(\sigma_2)\cdots \right]\nonumber \\
  && \;\;\;\;\times\left[ T_{a'b'c'd'}(\sigma_1) T_{a'e'f'g'}(\sigma_2)\cdots \right] 
\end{eqnarray}
gives the weight for the classical statistical mechanics problem.  It can be visualized as a ``bilayer'' network of two copies of the internal states defining the TNS (one each from the bra and the ket) and a single set of spins corresponding to the physical states.  Most importantly, the statistical weight is local and varies smoothly with the TNS tensors.  

Hence, we can rely on the well-known fact that in classical statistical mechanics with $d\geq 2$, a discrete symmetry breaking state is a stable {\em phase}.  Hence if a one point in parameter space exhibits symmetry breaking, for example the TNS which represents the actual ground state of the physical hamiltonian $H$, then generically the neighborhood of this point also exhibits broken symmetry.  Thus, unless the original TNS is finely tuned, arbitrary small variations of the tensor preserve broken symmetry.  This shows that it is {\em not} possible to construct a state with low energy density by exploring small variations of a TNS. Symmetry breaking is stable in this sense in $d\geq 2$ because the defects that destroy the order are {\em non-local}: domain walls with dimension $d-1$.  A large domain of linear size $L$ has a weight which is modified {\em locally} on of order $L^{d-1}$ tensors, and hence contributes a total weight of order $\epsilon^{L^{d-1}}$ to the wavefunction, which vanishes rapidly for large $L$.  Hence large domain walls are exponentially rare -- too rare to modify the long-range behavior of correlation functions.  

This strongly suggests that discrete broken symmetry order in $d\geq 2$ is variationally robust.  However, there is a loophole: there can be a variational state without broken symmetry which is {\em not} a small variation of the TNS representation of the ground state, but which nevertheless has low energy density.  We may imagine two possibilites.  First, there may be a TNS with a tensor which is not close to the ground state tensor, but which nevertheless has low energy.  This would appear to be an unlikely accident, and probably it is possible to argue that it only occurs with fine-tuning: with a small modification to the Hamiltonian, an accidental degeneracy like this can be split.  The second possibility is that there may be a state with low energy which does not have a TNS representation.

In fact, there definitely {\em are} low energy variational states which do not have the TNS form.  For example, we can consider a state with one-dimensional domain walls, inspired by the discussion in Secs.~\ref{sec:transv-field-ising},\ref{sec:doma-wall-interpr}.For concreteness take a square lattice.  We suppose the domain walls are rigid and infinitely long in the $y$ direction, and exist with probability amplitude $\epsilon$ on a given $x$ value.  Note that the latter assumption violates the locality of the TNS form, since the amplitude is not a product of local factors, which inevitably would be exponential in the length.  Nevertheless, one can still form a superposition of such domain walls.  Essentially this is the same as taking the MPS state in Eq.~\eqref{eq:2} and replacing each single spin $\sigma_i$ by an entire column of perfectly correlated spins in the $y$ direction at fixed $x$.  When $\epsilon$ is small, the average distance between domain walls is of order $1/\epsilon^2$, so that the energy density is order $\epsilon^2$.  

Clearly, however, this is not truly a disordered state.  It possesses long-range correlations along the $y$ direction.  It also violates the locality assumptions of the TNS, and consequently has other pathological properties.  For example, the entanglement entropy of a rectangular region of width $x$ and height $y$ scales only with $x$, a sub-area-law behavior.  We may reject this state as unphysical, or in any case readily diagnose such a state in numerics.  

We may imagine a variant of this state in which domain walls are introduced in both directions simultaneously, but still rigidly.  In this case the two-spin correlation function will decay exponentially in both the $x$ and $y$ directions.  However, the long-range order is not fully destroyed.  A four spin correlation function of the form
\begin{equation}
  \label{eq:26}
  C_{x,y} = \langle \sigma_{0,0} \sigma_{x,0} \sigma_{x,y} \sigma_{0,y}\rangle 
\end{equation}
will still not decay even for large $x$,$y$.  Hence this variant is not truly a disordered state.  

The above states with completely rigid domain walls are clearly unsatisfactory.  One may wonder, generally, whether there is some less artifical way to write a wavefunction which still realizes a state with fluctuating flipped domains of spins, but weights them differently than in a TNS, so that a finite weight is achieved for arbitrarily large domains.  We have not found a satisfactory general answer to this question.  States in which domain walls are localized are certainly possible, and simply represent {\em additional} symmetry breaking on top of the Ising order.  As argued in Sec.~\ref{sec:robustn-disord-stat}, this is always possible at low energy cost.  We suspect, without proof, that a weighting of arbitrarily large domains which restores all symmetries and yields a state which could be a generic ground state of some local Hamiltonian is not possible.  

In summary, the above arguments suggest that discrete symmetry breaking order {\em is} robust in $d\geq 2$, meaning that when a Hamiltonian exhibits discrete broken symmetry, any variational state which is a generic ground state (i.e. away from critical points) of some local Hamiltonian {\em and} which has this symmetry unbroken, must have a variational energy density which is strictly larger than that of the ground state.

\section{Topological order in two dimensions}
\label{sec:topol-order-two}

In this section, we consider the variational determination of topological order, in the same sense as discussed above for discrete symmetry breaking order.  We use topological order in the sense of Wen, to describe a fully gapped ground phase of matter at zero temperature, which exhibit ground state degeneracies in the thermodynamic limit on closed surfaces of non-trivial genus, and for which states within the degenerate subspace are indistinguishable by any local operator.  The essential physics of topological order is the existence of emergent anyonic excitations, with mutual statistics that cannot be obtained from any finite number of electrons.  Topological order exists only in dimensions greater than or equal to 2.  Even ignoring symmetry completely, the concept of topological order divides phases of matter into distinct topological classes.  A topological phase is locally stable to {\em any} local perturbation.

\subsection{Variational robustness of topological phases to states with "less" topological order}
\label{sec:robustn-topol-phas}

We first consider whether a topological phase is variationally robust to states with lower (or none) topological order. Despite the stability of the actual ground states to arbitrarily perturbations, we show that Hamiltonians for topological phases in two dimensions behave variationally like Hamiltonians for symmetry breaking order in one dimension.  That is, a wavefunction be found in a topological class {\em different} (lower, in a sense to be defined later) from that of the ground state, but with a variational energy density that can be made arbitrarily close to the exact value.  

Consider first the simplest example of a topological phase: the $Z_2$ or toric code phase.  It is exemplified by the model of Kitaev, which is exactly soluble:
\begin{equation}
  \label{eq:15}
  H_K = - \sum_p \prod_{i \in p} \sigma^z_i - \sum_s \prod_{i \in s} \sigma_i^x,
\end{equation}
where the spins live on the bonds $i$ of a 2d lattice, $p$ indicates plaquettes and $s$ indicates ``stars'', i.e. the set of bonds emanating from a given site.  The ground state can be considered as a uniform sum of all configurations in the $\sigma^x_i$ basis which satisfy the constraint that an even number of $\sigma^x_i=-1$ for every star.  From the structure of the exactly soluble wavefunction, one may directly verify the presence of topological order by for example calculating the quantized topological entanglement entropy, $S_{TEE} = \ln 2$, as defined in Refs.\onlinecite{levin2006detecting,kitaev2006topological}. 

To get more physical insight, one may represent the configurations by coloring the links with $\sigma^x_i=-1$. Then there must be an even number of colored links at every vertex.  This constraint corresponds to the existence of a conserved $Z_2$ ``electric'' flux, which can be defined as the number of colored links crossing a curve drawn on links of the dual lattice.  This flux can be odd or even, and is zero in the ground state for any contractible closed loop.  

Excited states may have a non-zero flux, which is due to an electric charge or {\sf e} particle inside the loop  (which have $\prod_{i\in s}\sigma^x_i = -1$ for some star $s$ inside the loop).  The {\sf e} particles are topological excitations of the $Z_2$ phase.  There are also ``magnetic'' {\sf m} particles, which correspond to defects in which the product of $\sigma_i^z$ around a closed loop is equal to $-1$ (and requires $\prod_{i \in p}\sigma^z_i =-1$ for some plaquette inside the loop).  The {\sf e} and {\sf m} particles are bosons, but if one considers ``mutual statistics'' together, they are relative semions: i.e. adiabatically transporting an {\sf e} particle around an {\sf m} particle incurs a change of the phase of the state by $\pi$. From these two particles one may also construct a composite {\sf e-m} particle, which is a fermion.  The {\sf e}, {\sf m}, and {\sf e-m} are the fundamental anyonic excitations of the $Z_2$ topological phase, and their existence may be regarded as the defining characteristic of the state.

The toric code phase is stable to {\em all} perturbations.  For example, the model on the square lattice has been extensively studied, and the topological phase has been shown to persist under the generic perturbations 
\begin{equation}
  \label{eq:16}
  H' = \sum_i h_z \sigma_i^z + h_x \sigma_i^x
\end{equation}
until the applied fields $h_x,h_z$ are of order one (beyond which a quantum phase transition occurs to a topologically trivial phase).  However, we may still ask whether it is possible to obtain a good approximation to the ground state energy density well within the topological phase by a topologically trivial wavefunction?

Following the logic of the previous section, we expect that the ground state within the $Z_2$ phase can be approximated to arbitrary accuracy by a tensor network state, or Projected Entangled Pair State (PEPS), which is the natural generalization of an MPS to higher dimensions.  Such a representation can be written explicitly for the exactly soluble Kitaev limit.  As for an MPS, any PEPS is the ground state of some local Hamiltonian (hence can be regarded as physical, and for example obeys the area law of entanglement entropy), and the variational energy {\em density} of a PEPS is a smooth function of the tensor components.  Hence if we can find an infinitesimal change of the tensor for a $Z_2$ state which destroys the topological order, we have found a variational state which satisfies the requirements.  In fact, such tensors have already been found by Chen {\em et al} in Ref.\onlinecite{chen2010tensor}. We do not repeat the details here as they are given very explicitly in the former paper.   Therefore it is indeeed possible to write a variational state which lacks topological order but has arbitrarily low energy density for the $Z_2$ topological phase.  

The mechanism behind the low energy variational wavefunctions is very similar to that discussed in the previous section.  A suitable deformation of the tensor defining the PEPS state introduces a low density (of order $\epsilon^2$ for an $O(\epsilon)$ deformation of the tensor) {\em free} ends into the ground state wavefunction.  Importantly, there are no long-range correlations between these ends in the variational state.  The presence of free ends removes the flux conservation on long distances and, as shown in Ref.\cite{chen2010tensor}, makes the topological entanglement entropy vanish. 

Physically, the free ends are electric {\sf e} particles, and the low density of these particles in the ground state corresponds to a {\em condensate} of electric charge.  We may then ask whether the quasiparticles of the toric code persist into this phase.  The {\sf e} particle no longer exists as a well-defined quasiparticle, like in any condensate, for which the particle number of the condensed particle becomes an uncertain variable.  The {\sf m} and {\sf e-m} particles, moreover, become {\em confined}, as their phase is scrambled by the delocalized background condensate of {\sf e} particles.  Hence none of the non-trivial anyons persist in the {\sf e} condensate state, so it is topologically trivial, consistent with vanishing topological entanglement entropy.  

The arbitrarily low energy density of the {\sf e} condensate state can be understood simply by the fact that the {\sf e} particle, being local, has a non-zero but finite energy, and so the energy density of a condensate can be rendered arbitrarily small but rendering the density of the {\sf e} particles low.  This is directly analogous to the low density soliton state which destroys symmetry breaking in one dimension.  

From this reasoning, we can immediately conclude that this is a general statement for topological phases in two dimensions: a state with ``less'' topological order can always be constructed by forming a low density condensation of an anyonic excitation of the topological phase.  The definition of ``less'' is in fact just those states which can be arrived at by anyon condensation. In general it may be possible for more than one such state to be constructed, if several bosonic anyons are available for condensation.  

\subsubsection{Half-integer spins}
\label{sec:half-integer-spins}

An interesting special case of the above discussion is the situation in which the system contains a half-integer spin per unit cell.  Then according to the generalized Lieb-Schulz-Mattis (gLSM) theorem, it is generally not possible to form a topological trivial state without broken symmetry or gapless excitations.\cite{hastings2004lieb,oshikawa2000commensurability} Suppose we have such a Hamiltonian whose ground state is a topological phase with no broken symmetry -- a topological spin liquid.  The arguments of this subsection still apply, so variational wavefunctions can be constructed without topological order and arbitrarily good energy density.  However, the gLSM argument implies that these variational states must either represent gapless phases or exhibit broken symmetry.  

Broken symmetry can arise very naturally in the states constructed by anyon condensation.  This occurs simply if the anyon which condenses carry symmetry quantum numbers.  For example, in $Z_2$ gauge theories of spin-1/2 spin liquids, the vison (or $m$ particle) carries space group quantum numbers, transforming under some projective representation of the symmetry group of the lattice.  The vison condensation then leads directly to valence bond order.  In general we expect that low energy density variational states that are topologically trivial and possess symmetry breaking order exist even when the true ground state is a featureless and topological spin liquid state.

\subsection{Robustness of trivial states to topological order}
\label{sec:robustn-triv-stat}

The converse question to the previous subsection is whether a ground state in a topologically trivial phase can be approximated well by one with topological order?  More generally we can ask whether a phase with a given topology can be approximated by one with ``more'' topological order -- but we will not attempt to answer this here. The analogous question in the case of symmetry breaking order was answered in Sec.~\ref{sec:robustn-disord-stat} with a resounding yes.  

However, an attempt to follow the argument of Sec.~\ref{sec:robustn-disord-stat} immediately runs into difficulty.  There, we showed how to construct a variational state by introducing a field coupling to the order parameter of the symmetry breaking.  However, for topological order, there is no local order parameter.  This construction fails immediately.  In general, since we expect that classes of topological order (including the trivial one) are {\em absolutely} stable, there can be no perturbation which defines a fiduciary Hamiltonian with topological order, when the ground state is trivial.  In passing, we note that if the original ground state is critical (i.e. $H$ is gapless), then perturbations that produce topological order may be possible.  However, gapless states are beyond the considerations of this paper.

We may turn to the TNS construction for guidance.  As shown in Ref.\cite{chen2010tensor}, tensors for states with $Z_2$ topological order must satisfy a symmetry requirement.  Any violation, however small, of this requirement destroys the topological order of the state.  Clearly, a generic state without $Z_2$ topological order has an asymmetric tensor.  As a finite object, there are no order of limits questions with respect to the tensor. It cannot be perturbed infinitesimally to restore its symmetry.  Hence at least within this construction it appears impossible to approximate a $Z_2$ topological phase by a trivial one.

A physical picture which confirms this notion comes from the string-net scheme of Levin and Wen\cite{levin2005string}.  They showed that topological phases, viewed approaching from a trivial state, arise by the proliferation/condensation of {\em infinitely long} extended strings or string-nets.  Clearly, if the ground state of our Hamiltonian A is in the trivial phase, the strings must be finite there, i.e. there is a non-zero string tension, or energy per length of string.  Any variational state with infinitely long strings must pay this string tension, which is a volume energy since the strings are dense in the topological phase.  The situation seems analogous to the destruction of symmetry breaking order, which requires condensation of domain walls of dimension $d-1>0$ in $d\geq 2$.   Hence we expect that topologically trivial phases are variationally robust to topological ones.

\section{Discussion}
\label{sec:discussion}

We have discussed the question of variational robustness of certain phases against others: can a Hamiltonian with a ground state in phase A be approximated by a physical wavefunction in another phase B, with arbitrarily small energy density?  In fact, in surprisingly many cases, the answer is yes.  Two cases in which it is not obviously possible are (1) A is a symmetry broken state and the symmetry is unbroken in B, in dimensions two and larger, and (2) A is a trivial state and B has topological order.  In case (1), the restriction that the symmetric state B be physical is non-trivial and crucial.  

The approach of this paper has been to construct examples of principle, to show that in many cases an approximation {\em is} possible, i.e. that phase A is {\em not} variationally robust to phase B.  The examples were based on a simple paradigm of defect proliferation.  This does not mean this is the only way in which a low energy variational state may manifest.  Rather, we intend the construction here as a proof of principle, to show that in these cases the finding of a low energy density variational state does not necessarily mean the ground state phase has been properly identified.  

Probably the major context in which these results may be relevant is the study of spin liquid phases of quantum magnets, where variational methods are a dominant approach.  The fact that topological spin liquids are not variationally robust to trivial states suggests that variational methods tend to overestimate the dominance of topologically trivial but ordered phases.  Conversely, the variational robustness of trivial phases to topological order means that a good variational state which has topological order {\em is} a strong argument for topological order in the ground state.  It is important to note that the arguments expressed in this paper presuppose the existence of an excitation gap, and this must be checked independently in a numerical calculation.

In practice, there are many means beyond just inspecting the energy to evaluate variational states.  Many of the low energy variational states constructed in this paper have emergent long length scales, which could be detected by various measurements.  Correct properties of the ground state might be obtained from a careful study of the wavefunction on shorter scales.  

One may envision pursuing these ideas further.  Quite likely some of the statements in this paper could be made mathematically rigorous.  The yes/no question of variational robustness formulated here is only crudest type of statement one might make about variational states.  It would be highly desirable to be more quantitative about various physical properties.  For example, one would like to know how much correlation functions or reduced density matrices over regions of a given size can differ between the ground and variational states, given a particular difference in energy density of the two states.  

{\em Acknowledgements.---} I thank Lucile Savary and Tarun Grover for discussions.  This research was supported by the NSF through grant DMR-12-06809.  

\bibliography{var.bib}

%merlin.mbs apsrev4-1.bst 2010-07-25 4.21a (PWD, AO, DPC) hacked
%Control: key (0)
%Control: author (8) initials jnrlst
%Control: editor formatted (1) identically to author
%Control: production of article title (-1) disabled
%Control: page (0) single
%Control: year (1) truncated
%Control: production of eprint (0) enabled
\begin{thebibliography}{16}%
\makeatletter
\providecommand \@ifxundefined [1]{%
 \@ifx{#1\undefined}
}%
\providecommand \@ifnum [1]{%
 \ifnum #1\expandafter \@firstoftwo
 \else \expandafter \@secondoftwo
 \fi
}%
\providecommand \@ifx [1]{%
 \ifx #1\expandafter \@firstoftwo
 \else \expandafter \@secondoftwo
 \fi
}%
\providecommand \natexlab [1]{#1}%
\providecommand \enquote  [1]{``#1''}%
\providecommand \bibnamefont  [1]{#1}%
\providecommand \bibfnamefont [1]{#1}%
\providecommand \citenamefont [1]{#1}%
\providecommand \href@noop [0]{\@secondoftwo}%
\providecommand \href [0]{\begingroup \@sanitize@url \@href}%
\providecommand \@href[1]{\@@startlink{#1}\@@href}%
\providecommand \@@href[1]{\endgroup#1\@@endlink}%
\providecommand \@sanitize@url [0]{\catcode `\\12\catcode `\$12\catcode
  `\&12\catcode `\#12\catcode `\^12\catcode `\_12\catcode `\%12\relax}%
\providecommand \@@startlink[1]{}%
\providecommand \@@endlink[0]{}%
\providecommand \url  [0]{\begingroup\@sanitize@url \@url }%
\providecommand \@url [1]{\endgroup\@href {#1}{\urlprefix }}%
\providecommand \urlprefix  [0]{URL }%
\providecommand \Eprint [0]{\href }%
\providecommand \doibase [0]{http://dx.doi.org/}%
\providecommand \selectlanguage [0]{\@gobble}%
\providecommand \bibinfo  [0]{\@secondoftwo}%
\providecommand \bibfield  [0]{\@secondoftwo}%
\providecommand \translation [1]{[#1]}%
\providecommand \BibitemOpen [0]{}%
\providecommand \bibitemStop [0]{}%
\providecommand \bibitemNoStop [0]{.\EOS\space}%
\providecommand \EOS [0]{\spacefactor3000\relax}%
\providecommand \BibitemShut  [1]{\csname bibitem#1\endcsname}%
\let\auto@bib@innerbib\@empty
%</preamble>
\bibitem [{\citenamefont {Wen}(2002)}]{wen2002quantum}%
  \BibitemOpen
  \bibfield  {author} {\bibinfo {author} {\bibfnamefont {X.-G.}\ \bibnamefont
  {Wen}},\ }\href@noop {} {\bibfield  {journal} {\bibinfo  {journal} {Physical
  Review B}\ }\textbf {\bibinfo {volume} {65}},\ \bibinfo {pages} {165113}
  (\bibinfo {year} {2002})}\BibitemShut {NoStop}%
\bibitem [{\citenamefont {Iqbal}\ \emph {et~al.}(2013)\citenamefont {Iqbal},
  \citenamefont {Becca}, \citenamefont {Sorella},\ and\ \citenamefont
  {Poilblanc}}]{iqbal2013gapless}%
  \BibitemOpen
  \bibfield  {author} {\bibinfo {author} {\bibfnamefont {Y.}~\bibnamefont
  {Iqbal}}, \bibinfo {author} {\bibfnamefont {F.}~\bibnamefont {Becca}},
  \bibinfo {author} {\bibfnamefont {S.}~\bibnamefont {Sorella}}, \ and\
  \bibinfo {author} {\bibfnamefont {D.}~\bibnamefont {Poilblanc}},\ }\href@noop
  {} {\bibfield  {journal} {\bibinfo  {journal} {Physical Review B}\ }\textbf
  {\bibinfo {volume} {87}},\ \bibinfo {pages} {060405} (\bibinfo {year}
  {2013})}\BibitemShut {NoStop}%
\bibitem [{\citenamefont {Lu}\ \emph {et~al.}(2011)\citenamefont {Lu},
  \citenamefont {Ran},\ and\ \citenamefont {Lee}}]{lu2011z}%
  \BibitemOpen
  \bibfield  {author} {\bibinfo {author} {\bibfnamefont {Y.-M.}\ \bibnamefont
  {Lu}}, \bibinfo {author} {\bibfnamefont {Y.}~\bibnamefont {Ran}}, \ and\
  \bibinfo {author} {\bibfnamefont {P.~A.}\ \bibnamefont {Lee}},\ }\href@noop
  {} {\bibfield  {journal} {\bibinfo  {journal} {Physical Review B}\ }\textbf
  {\bibinfo {volume} {83}},\ \bibinfo {pages} {224413} (\bibinfo {year}
  {2011})}\BibitemShut {NoStop}%
\bibitem [{\citenamefont {Clark}\ \emph {et~al.}(2011)\citenamefont {Clark},
  \citenamefont {Abanin},\ and\ \citenamefont {Sondhi}}]{clark2011nature}%
  \BibitemOpen
  \bibfield  {author} {\bibinfo {author} {\bibfnamefont {B.}~\bibnamefont
  {Clark}}, \bibinfo {author} {\bibfnamefont {D.}~\bibnamefont {Abanin}}, \
  and\ \bibinfo {author} {\bibfnamefont {S.}~\bibnamefont {Sondhi}},\
  }\href@noop {} {\bibfield  {journal} {\bibinfo  {journal} {Physical review
  letters}\ }\textbf {\bibinfo {volume} {107}},\ \bibinfo {pages} {087204}
  (\bibinfo {year} {2011})}\BibitemShut {NoStop}%
\bibitem [{\citenamefont {Hermele}\ \emph {et~al.}(2008)\citenamefont
  {Hermele}, \citenamefont {Ran}, \citenamefont {Lee},\ and\ \citenamefont
  {Wen}}]{hermele2008properties}%
  \BibitemOpen
  \bibfield  {author} {\bibinfo {author} {\bibfnamefont {M.}~\bibnamefont
  {Hermele}}, \bibinfo {author} {\bibfnamefont {Y.}~\bibnamefont {Ran}},
  \bibinfo {author} {\bibfnamefont {P.~A.}\ \bibnamefont {Lee}}, \ and\
  \bibinfo {author} {\bibfnamefont {X.-G.}\ \bibnamefont {Wen}},\ }\href@noop
  {} {\bibfield  {journal} {\bibinfo  {journal} {Physical Review B}\ }\textbf
  {\bibinfo {volume} {77}},\ \bibinfo {pages} {224413} (\bibinfo {year}
  {2008})}\BibitemShut {NoStop}%
\bibitem [{\citenamefont {Zhou}\ \emph {et~al.}(2008)\citenamefont {Zhou},
  \citenamefont {Lee}, \citenamefont {Ng},\ and\ \citenamefont
  {Zhang}}]{zhou20084}%
  \BibitemOpen
  \bibfield  {author} {\bibinfo {author} {\bibfnamefont {Y.}~\bibnamefont
  {Zhou}}, \bibinfo {author} {\bibfnamefont {P.~A.}\ \bibnamefont {Lee}},
  \bibinfo {author} {\bibfnamefont {T.-K.}\ \bibnamefont {Ng}}, \ and\ \bibinfo
  {author} {\bibfnamefont {F.-C.}\ \bibnamefont {Zhang}},\ }\href@noop {}
  {\bibfield  {journal} {\bibinfo  {journal} {Physical review letters}\
  }\textbf {\bibinfo {volume} {101}},\ \bibinfo {pages} {197201} (\bibinfo
  {year} {2008})}\BibitemShut {NoStop}%
\bibitem [{\citenamefont {Lawler}\ \emph {et~al.}(2008)\citenamefont {Lawler},
  \citenamefont {Paramekanti}, \citenamefont {Kim},\ and\ \citenamefont
  {Balents}}]{lawler2008gapless}%
  \BibitemOpen
  \bibfield  {author} {\bibinfo {author} {\bibfnamefont {M.~J.}\ \bibnamefont
  {Lawler}}, \bibinfo {author} {\bibfnamefont {A.}~\bibnamefont {Paramekanti}},
  \bibinfo {author} {\bibfnamefont {Y.~B.}\ \bibnamefont {Kim}}, \ and\
  \bibinfo {author} {\bibfnamefont {L.}~\bibnamefont {Balents}},\ }\href@noop
  {} {\bibfield  {journal} {\bibinfo  {journal} {Physical review letters}\
  }\textbf {\bibinfo {volume} {101}},\ \bibinfo {pages} {197202} (\bibinfo
  {year} {2008})}\BibitemShut {NoStop}%
\bibitem [{\citenamefont {Yunoki}\ and\ \citenamefont
  {Sorella}(2006)}]{yunoki2006two}%
  \BibitemOpen
  \bibfield  {author} {\bibinfo {author} {\bibfnamefont {S.}~\bibnamefont
  {Yunoki}}\ and\ \bibinfo {author} {\bibfnamefont {S.}~\bibnamefont
  {Sorella}},\ }\href@noop {} {\bibfield  {journal} {\bibinfo  {journal}
  {Physical Review B}\ }\textbf {\bibinfo {volume} {74}},\ \bibinfo {pages}
  {014408} (\bibinfo {year} {2006})}\BibitemShut {NoStop}%
\bibitem [{\citenamefont {Motrunich}(2005)}]{motrunich2005variational}%
  \BibitemOpen
  \bibfield  {author} {\bibinfo {author} {\bibfnamefont {O.~I.}\ \bibnamefont
  {Motrunich}},\ }\href@noop {} {\bibfield  {journal} {\bibinfo  {journal}
  {Physical Review B}\ }\textbf {\bibinfo {volume} {72}},\ \bibinfo {pages}
  {045105} (\bibinfo {year} {2005})}\BibitemShut {NoStop}%
\bibitem [{\citenamefont {Ran}\ \emph {et~al.}(2007)\citenamefont {Ran},
  \citenamefont {Hermele}, \citenamefont {Lee},\ and\ \citenamefont
  {Wen}}]{ran2007projected}%
  \BibitemOpen
  \bibfield  {author} {\bibinfo {author} {\bibfnamefont {Y.}~\bibnamefont
  {Ran}}, \bibinfo {author} {\bibfnamefont {M.}~\bibnamefont {Hermele}},
  \bibinfo {author} {\bibfnamefont {P.~A.}\ \bibnamefont {Lee}}, \ and\
  \bibinfo {author} {\bibfnamefont {X.-G.}\ \bibnamefont {Wen}},\ }\href@noop
  {} {\bibfield  {journal} {\bibinfo  {journal} {Physical review letters}\
  }\textbf {\bibinfo {volume} {98}},\ \bibinfo {pages} {117205} (\bibinfo
  {year} {2007})}\BibitemShut {NoStop}%
\bibitem [{\citenamefont {Levin}\ and\ \citenamefont
  {Wen}(2006)}]{levin2006detecting}%
  \BibitemOpen
  \bibfield  {author} {\bibinfo {author} {\bibfnamefont {M.}~\bibnamefont
  {Levin}}\ and\ \bibinfo {author} {\bibfnamefont {X.-G.}\ \bibnamefont
  {Wen}},\ }\href@noop {} {\bibfield  {journal} {\bibinfo  {journal} {Physical
  review letters}\ }\textbf {\bibinfo {volume} {96}},\ \bibinfo {pages}
  {110405} (\bibinfo {year} {2006})}\BibitemShut {NoStop}%
\bibitem [{\citenamefont {Kitaev}\ and\ \citenamefont
  {Preskill}(2006)}]{kitaev2006topological}%
  \BibitemOpen
  \bibfield  {author} {\bibinfo {author} {\bibfnamefont {A.}~\bibnamefont
  {Kitaev}}\ and\ \bibinfo {author} {\bibfnamefont {J.}~\bibnamefont
  {Preskill}},\ }\href@noop {} {\bibfield  {journal} {\bibinfo  {journal}
  {Physical review letters}\ }\textbf {\bibinfo {volume} {96}},\ \bibinfo
  {pages} {110404} (\bibinfo {year} {2006})}\BibitemShut {NoStop}%
\bibitem [{\citenamefont {Chen}\ \emph {et~al.}(2010)\citenamefont {Chen},
  \citenamefont {Zeng}, \citenamefont {Gu}, \citenamefont {Chuang},\ and\
  \citenamefont {Wen}}]{chen2010tensor}%
  \BibitemOpen
  \bibfield  {author} {\bibinfo {author} {\bibfnamefont {X.}~\bibnamefont
  {Chen}}, \bibinfo {author} {\bibfnamefont {B.}~\bibnamefont {Zeng}}, \bibinfo
  {author} {\bibfnamefont {Z.-C.}\ \bibnamefont {Gu}}, \bibinfo {author}
  {\bibfnamefont {I.~L.}\ \bibnamefont {Chuang}}, \ and\ \bibinfo {author}
  {\bibfnamefont {X.-G.}\ \bibnamefont {Wen}},\ }\href@noop {} {\bibfield
  {journal} {\bibinfo  {journal} {Physical Review B}\ }\textbf {\bibinfo
  {volume} {82}},\ \bibinfo {pages} {165119} (\bibinfo {year}
  {2010})}\BibitemShut {NoStop}%
\bibitem [{\citenamefont {Hastings}(2004)}]{hastings2004lieb}%
  \BibitemOpen
  \bibfield  {author} {\bibinfo {author} {\bibfnamefont {M.~B.}\ \bibnamefont
  {Hastings}},\ }\href@noop {} {\bibfield  {journal} {\bibinfo  {journal}
  {Physical Review B}\ }\textbf {\bibinfo {volume} {69}},\ \bibinfo {pages}
  {104431} (\bibinfo {year} {2004})}\BibitemShut {NoStop}%
\bibitem [{\citenamefont {Oshikawa}(2000)}]{oshikawa2000commensurability}%
  \BibitemOpen
  \bibfield  {author} {\bibinfo {author} {\bibfnamefont {M.}~\bibnamefont
  {Oshikawa}},\ }\href@noop {} {\bibfield  {journal} {\bibinfo  {journal}
  {Physical review letters}\ }\textbf {\bibinfo {volume} {84}},\ \bibinfo
  {pages} {1535} (\bibinfo {year} {2000})}\BibitemShut {NoStop}%
\bibitem [{\citenamefont {Levin}\ and\ \citenamefont
  {Wen}(2005)}]{levin2005string}%
  \BibitemOpen
  \bibfield  {author} {\bibinfo {author} {\bibfnamefont {M.~A.}\ \bibnamefont
  {Levin}}\ and\ \bibinfo {author} {\bibfnamefont {X.-G.}\ \bibnamefont
  {Wen}},\ }\href@noop {} {\bibfield  {journal} {\bibinfo  {journal} {Physical
  Review B}\ }\textbf {\bibinfo {volume} {71}},\ \bibinfo {pages} {045110}
  (\bibinfo {year} {2005})}\BibitemShut {NoStop}%
\end{thebibliography}%

\end{document}